\pdfoutput=1

\documentclass[12pt,a4paper]{article}

\usepackage{ifthen}
\usepackage{etoolbox}
\usepackage{forarray}

\newboolean{articletitles}
\setboolean{articletitles}{true}

\newboolean{inbibliography}
\setboolean{inbibliography}{false}

\usepackage[american]{babel}

\usepackage{microtype}
\usepackage{xspace} 

\usepackage{caption}

\usepackage{graphicx} 
\usepackage{color}
\usepackage{colortbl}

\usepackage{booktabs}
\usepackage{longtable}
\usepackage{multirow}

\usepackage{siunitx}
\DeclareSIUnit\clight{\ensuremath{\mathit{c}}}

\usepackage{amsmath} 
\usepackage{amssymb}
\usepackage{amsfonts}
\usepackage{upgreek} 
\usepackage{fancyvrb} 
\usepackage{xfrac}
\usepackage{csquotes}

\usepackage{hyperref}
\usepackage[all]{hypcap} 

\usepackage{cleveref}
\crefname{chapter}{Ch.\@}{Chs.\@}
\crefname{section}{Sec.\@}{Secs.\@}
\crefname{subsection}{Sec.\@}{Secs.\@}
\crefname{appendix}{Appendix\@}{Appendices\@}
\crefname{figure}{Fig.\@}{Figures\@}
\crefname{table}{Table\@}{Tables\@}
\crefname{equation}{Eq.\@}{Eqs.\@}

\usepackage{cite}
\usepackage{mciteplus}

\begin{document}

\renewcommand{\thefootnote}{\fnsymbol{footnote}}
\setcounter{footnote}{1}

%!TEX root = ../main.tex

\begin{titlepage}
\pagenumbering{roman}

% Header ---------------------------------------------------
\vspace*{-1.5cm}
\centerline{\large EUROPEAN ORGANIZATION FOR NUCLEAR RESEARCH (CERN)}
\vspace*{1.5cm}
\hspace*{-0.5cm}
\begin{tabular*}{\linewidth}{lc@{\extracolsep{\fill}}r}
\vspace*{-2.7cm}\mbox{\!\!\!\includegraphics[width=.14\textwidth]{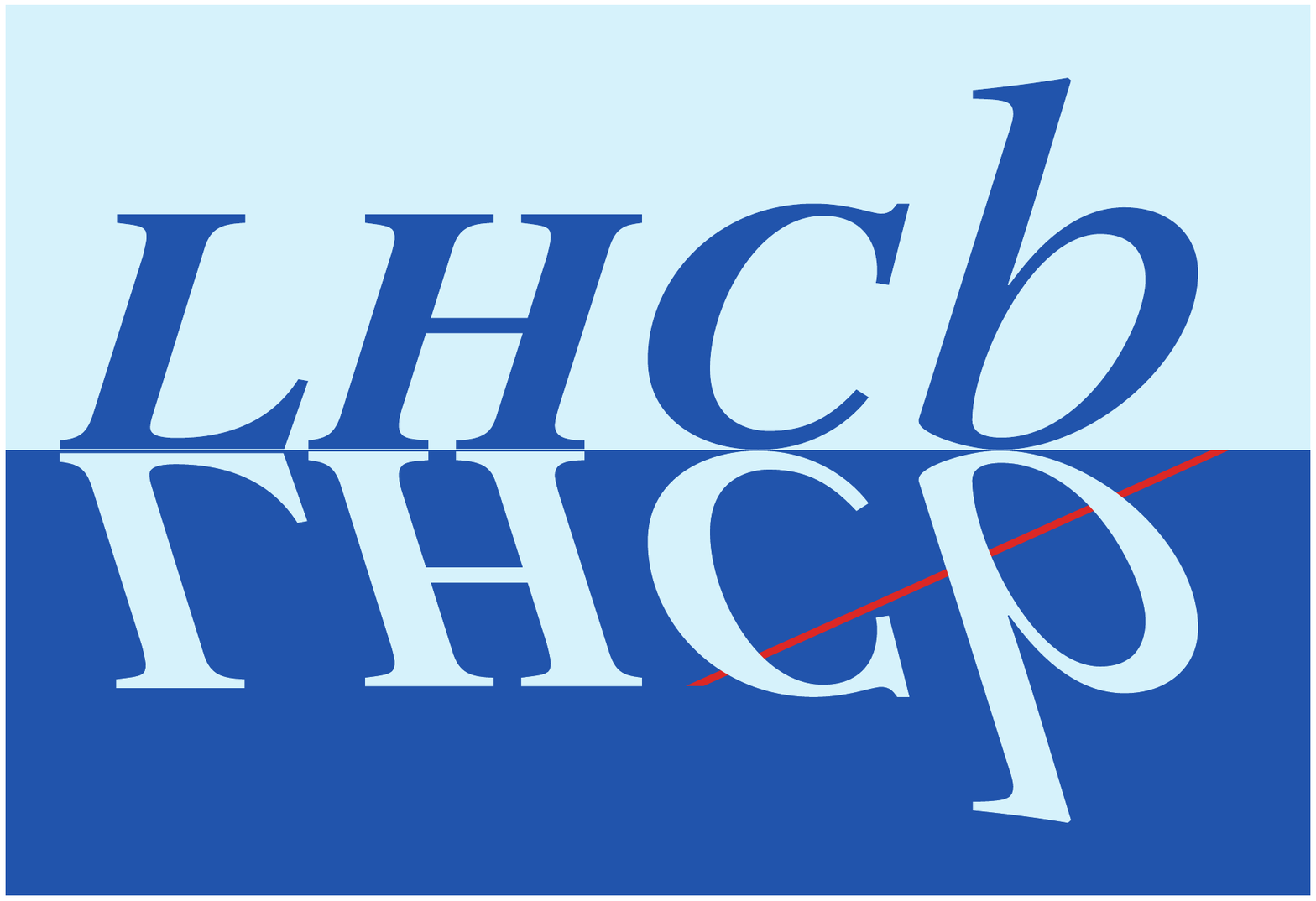}} & & \\
 & & CERN-PH-EP-2015-076 \\  % ID 
 & & LHCb-PAPER-2015-004 \\  % ID 
 & & March 24, 2015 \\ % Date 
 & & \\%Version 6.0\\
 & & \\
 & & \\
\end{tabular*}

\vspace*{3.5cm}

% Title --------------------------------------------------
{\bf\boldmath\huge
\begin{center}
  Measurement of $C\!P$ violation in $B^0 \!\rightarrow {J\mskip -3mu/\mskip -2mu\psi\mskip 2mu} {K^0_{\rm\scriptscriptstyle S}}$ decays
\end{center}
}

\vspace*{0.5cm}

% Authors -------------------------------------------------
\begin{center}
The LHCb collaboration\footnote{Authors are listed at the end of this Letter.}
\end{center}

\vspace*{1.5cm}

% Abstract -----------------------------------------------
\begin{abstract}
\noindent
Measurements are presented of the $C\!P$ violation observables $S$ and $C$ in
the decays of $B^0$ and ${\kern 0.18em\overline{\kern -0.18em B}}{}^0$ mesons to
the ${J\mskip -3mu/\mskip -2mu\psi\mskip 2mu} {K^0_{\rm\scriptscriptstyle S}}$
final state. The data sample corresponds to an integrated luminosity of
$\SI{3.0}{fb^{-1}}$ collected with the LHCb experiment in proton-proton
collisions at center-of-mass energies of $\num{7}$ and $\SI{8}{\tera\eV}$.
The analysis of the time evolution of $\num{41500}$ $B^0$ and ${\kern
0.18em\overline{\kern -0.18em B}}{}^0$ decays yields \mbox{$S = 0.731 \pm 0.035 \,
  \text{(stat)} \pm 0.020 \,\text{(syst)}$} and \mbox{$C = -0.038 \pm 0.032 \,
    \text{(stat)} \pm 0.005\,\text{(syst)}$}. In the Standard Model, $S$ equals
$\sin(2\beta)$ to a good level of precision. The values are consistent with the
current world averages and with the Standard Model expectations.
\end{abstract}

\vspace*{1.5cm}

\begin{center}
  Submitted to Phys.~Rev.~Lett.
\end{center}

\vspace{\fill}

{\footnotesize 
\centerline{\copyright~CERN on behalf of the LHCb collaboration, license \href{http://creativecommons.org/licenses/by/4.0/}{CC-BY-4.0}.}}
\vspace*{2mm}

\end{titlepage}

\newpage
\setcounter{page}{2}
\mbox{~}
\cleardoublepage

\renewcommand{\thefootnote}{\arabic{footnote}}
\setcounter{footnote}{0}

\pagestyle{plain}
\setcounter{page}{1}
\pagenumbering{arabic}

%!TEX root = ../main.tex

The violation of charge-parity ($C\!P$) conservation in processes involving $B$
mesons was first observed in the \enquote{golden mode} $B^0 \!\rightarrow {J\mskip
-3mu/\mskip -2mu\psi\mskip 2mu} {K^0_{\rm\scriptscriptstyle S}}$ by the BaBar
and Belle experiments at the asymmetric $e^{+}e^{-}$ colliders
PEP-II and KEKB~\cite{sin2beta:babar:firstobs,sin2beta:belle:firstobs}. Since
then, measurements of $C\!P$ violation in this decay mode have reached a
precision at the level of $10^{-2}$~\cite{sin2beta:babar,sin2beta:belle}. Thus,
these measurements play an important role in constraining and testing the 
quark-flavor sector of the Standard Model~\cite{Carter:1980tk,Bigi:1981qs}, which
relates $C\!P$-violating observables to a single irreducible phase in the
Cabibbo-Kobayashi-Maskawa (CKM) quark-mixing
matrix~\cite{Cabibbo,KobayashiMaskawa}. As the ${J\mskip -3mu/\mskip
-2mu\psi\mskip 2mu} {K^0_{\rm\scriptscriptstyle S}}$ final state is common to
both the $B^0$ and the ${\kern 0.18em\overline{\kern -0.18em B}}{}^0$ meson
decays, the interference between the amplitudes for the direct decay and for the
decay after $B^0$--${\kern 0.18em\overline{\kern -0.18em B}}{}^0$ oscillation
results in a decay-time dependent $C\!P$ asymmetry between the time-dependent
decay rates of $B^0$ and ${\kern 0.18em\overline{\kern -0.18em B}}{}^0$
mesons,
\begin{equation}
  \mathcal{A}(t) 
   \equiv \frac{\Gamma({\kern 0.18em\overline{\kern -0.18em B}}{}^0(t)\!\rightarrow{J\mskip -3mu/\mskip -2mu\psi\mskip 2mu}{K^0_{\rm\scriptscriptstyle S}}) - \Gamma(B^0(t)\!\rightarrow{J\mskip -3mu/\mskip -2mu\psi\mskip 2mu}{K^0_{\rm\scriptscriptstyle S}})}
                {\Gamma({\kern 0.18em\overline{\kern -0.18em B}}{}^0(t)\!\rightarrow{J\mskip -3mu/\mskip -2mu\psi\mskip 2mu}{K^0_{\rm\scriptscriptstyle S}}) + \Gamma(B^0(t)\!\rightarrow{J\mskip -3mu/\mskip -2mu\psi\mskip 2mu}{K^0_{\rm\scriptscriptstyle S}})} 
   =      \frac{S \sin(\mathrm{\Delta} m\, t) - C \cos(\mathrm{\Delta} m\, t)}
                {\cosh(\frac{\mathrm{\Delta} \Gamma\, t}{2}) + A_{\mathrm{\Delta} \Gamma}\sinh(\frac{\mathrm{\Delta} \Gamma\, t}{2})}.
\end{equation}
Here, $B^0(t)$ and ${\kern 0.18em\overline{\kern -0.18em B}}{}^0(t)$ indicate
the flavor of the $B$ meson at production, while $t$ indicates the decay time.
The parameters $\mathrm{\Delta} m$ and $\mathrm{\Delta} \Gamma$ are the mass and
the decay width differences between the heavy and light mass eigenstates of the
$B^0$--${\kern 0.18em\overline{\kern -0.18em B}}{}^0$ system, and $S$, $C$,
and $A_{\mathrm{\Delta} \Gamma}$ are $C\!P$ observables. As $\mathrm{\Delta}
\Gamma$ is negligible for the $B^0$--${\kern 0.18em\overline{\kern -0.18em
B}}{}^0$ system~\cite{HFAG}, the time-dependent asymmetry simplifies to
$\mathcal{A}(t) = S\sin(\mathrm{\Delta} m\, t) - C \cos(\mathrm{\Delta} m\, t)$.

The $B^0 \!\rightarrow {J\mskip -3mu/\mskip -2mu\psi\mskip 2mu}
{K^0_{\rm\scriptscriptstyle S}}$ decay is dominated by a
$\overline{b}\!\rightarrow c\overline{c}\,\overline{s}$
transition,\footnote{Mention of a particular decay mode implies the inclusion of
charge-conjugate states except when the measurement of $C\!P$ violation is
involved.} and $C\!P$ violation in the decay is expected to be negligible at the
current level of experimental precision, giving $C \approx 0$. This allows to
identify $S$ with $\sin(2\beta)$, where $\beta\equiv\arg[-(V_{cd}^{\vphantom{\ast}} V_{cb}^{\ast})
/(V_{td}^{\vphantom{\ast}} V_{tb}^{\ast})]$ is one of the angles of the CKM
triangle. Other measurements that constrain this triangle predict $\sin(2\beta)$
as $0.771 \pm^{0.017}_{0.041}$~\cite{Charles:2015gya}, giving a small
discrepancy with respect to the average of direct measurements,
$0.682\pm0.019$~\cite{HFAG}, where the most precise input comes from a $C\!P$
violation measurement in $B^0 \!\rightarrow {J\mskip -3mu/\mskip -2mu\psi\mskip
2mu} {K^0_{\rm\scriptscriptstyle S}}$ decays by the Belle experiment, $S=0.670
\pm 0.029\,\text{(stat)}\pm 0.013\,\text{(syst)}$~\cite{sin2beta:belle}. To
clarify the CKM picture, both better experimental precision and improved
understanding of higher-order contributions to the decay amplitudes are
required~\cite{CPVBs2JpsiKS:lhcb,*Aaij:2014vda,DeBruyn:2014oga,*Frings:2015eva}.

The analysis presented in this Letter supersedes a previous measurement by
LHCb~\cite{sin2beta:lhcb:1fb}, which was performed on data corresponding to an
integrated luminosity of $\SI{1.0}{\per\femto\barn}$ at a center-of-mass energy
of $\SI{7}{TeV}$. By adding data corresponding to $\SI{2}{\per\femto\barn}$ at
$\SI{8}{TeV}$ and using an optimized selection and additional \enquote{flavor
tagging} algorithms to identify the quark content of the $B$ meson at
production, we increase the statistical power of the analysis by almost a factor
$\num{6}$.

The LHCb detector~\cite{Alves:2008zz,LHCb-DP-2014-002} is a single-arm forward
spectrometer covering the pseudorapidity range $2<\eta <5$, designed for the
study of particles containing $b$ or $c$ quarks. The detector includes a 
high-precision tracking system consisting of a silicon-strip vertex detector
surrounding the $pp$ interaction region, a large-area silicon-strip detector
located upstream of a dipole magnet, and three stations of silicon-strip
detectors and straw drift tubes placed downstream of the magnet. Different types
of charged hadrons are distinguished using information from two ring-imaging
Cherenkov detectors. Photon, electron, and hadron candidates are identified by a
calorimeter system consisting of scintillating-pad and preshower detectors, an
electromagnetic calorimeter, and a hadronic calorimeter. Muons are identified by
a system composed of alternating layers of iron and multiwire proportional
chambers. The online event selection system (trigger)~\cite{Aaij:2012me}
consists of a hardware stage, based on information from the calorimeter and muon
systems, followed by a software stage.

The analysis is performed with $B^0 \!\rightarrow {J\mskip -3mu/\mskip
-2mu\psi\mskip 2mu} {K^0_{\rm\scriptscriptstyle S}}$ candidates reconstructed in
the ${J\mskip -3mu/\mskip -2mu\psi\mskip 2mu} \!\rightarrow \mu^{+}\mu^{-}$ and
${K^0_{\rm\scriptscriptstyle S}} \!\rightarrow \pi^{+}\pi^{-}$ final states. Two
oppositely charged particles, identified as muons with high momentum and high
transverse momentum, are required to originate from a common space-point
(vertex) and to have an invariant mass in a range
$\SI[per-mode=symbol]{\pm60}{\mega\eV\per\square\clight}$ around the known ${J\mskip
-3mu/\mskip -2mu\psi\mskip 2mu}$ mass~\cite{PDG2014}. Since the $B^0$ meson has
a lifetime of $\SI{1.5}{\pico\second}$ and has high momentum, the resulting
${J\mskip -3mu/\mskip -2mu\psi\mskip 2mu}$ candidate is required to be
significantly separated from all reconstructed $pp$ collision points (primary
vertices (PVs)). The ${K^0_{\rm\scriptscriptstyle S}}$ candidates are formed
from two oppositely charged, high-momentum pion candidates with a clear
separation from any PV in the event. Candidates decaying early enough for the
final-state pions to be reconstructed in the vertex detector are characterized
as \emph{long} candidates and are required to have an invariant mass within
$\SI[per-mode=symbol]{\pm15}{\mega\eV\per\square\clight}$ of the known
${K^0_{\rm\scriptscriptstyle S}}$ mass~\cite{PDG2014}. The
${K^0_{\rm\scriptscriptstyle S}}$ candidates that decay later, such that track
segments of the pions cannot be formed in the vertex detector, are called
\emph{downstream} candidates; these have a larger momentum resolution than the
long candidates, and thus the corresponding $\pi^{+}\pi^{-}$ pairs are required
to have an invariant mass within $\SI[per-mode=symbol]{\pm55}{\mega\eV\per\square\clight}$
of the known ${K^0_{\rm\scriptscriptstyle S}}$ mass. A good vertex fit quality
and sufficient separation from the $B^0$ decay vertex are required for the
${K^0_{\rm\scriptscriptstyle S}}$ candidate's decay vertex. To eliminate
background contributions from $\varLambda^0_{b} \!\rightarrow {J\mskip -3mu/\mskip-
2mu\psi\mskip 2mu}\varLambda$ decays, the $\pi^{+}$ ($\pi^{-}$) candidate has to
fulfill particle identification requirements if the invariant mass under a
$p\pi^{-}$ ($\pi^{+}\overline{p}$) mass hypothesis is compatible with the
$\varLambda$ mass.

The $B^0$ candidates are reconstructed from ${J\mskip -3mu/\mskip -2mu\psi\mskip
2mu}$ and ${K^0_{\rm\scriptscriptstyle S}}$ candidates that form a good quality
vertex. Their decay time $t$ is obtained from a fit to the full decay chain
while constraining the production vertex of the $B^0$ to the associated
PV~\cite{Hulsbergen:2005pu}. The reconstructed $B^0$ candidate mass $m$ is
obtained from a similar fit with the $\mu^{+}\mu^{-}$ and $\pi^{+}\pi^{-}$
invariant masses constrained to the known ${J\mskip -3mu/\mskip -2mu\psi\mskip
2mu}$ and ${K^0_{\rm\scriptscriptstyle S}}$ masses. The latter fit must satisfy
loose requirements on its quality, and resulting candidates are only retained if
$5230 < m < \SI[per-mode=symbol]{5330}{\mega\eV\per\square\clight}$ and
$0.3<t<\SI{18.3}{\pico\second}$. The fit uncertainty $\sigma_{t}$ on the decay
time is required to be smaller than $\SI{200}{\femto\second}$, which is well
above the average resolution of $\SI{55}{\femto\second}$
($\SI{65}{\femto\second}$) for candidates with long (downstream)
${K^0_{\rm\scriptscriptstyle S}}$ daughters. The quantity $\sigma_{t}$ is used
later in the analysis as an estimate of the per-candidate decay-time resolution.
Multiple PVs and, in a small fraction of events, multiple $B^0 \!\rightarrow
{J\mskip -3mu/\mskip -2mu\psi\mskip 2mu} {K^0_{\rm\scriptscriptstyle S}}$
candidates, lead to multiple ($B^{0},\text{PV}$) pairs per event. In such cases,
one pair is chosen at random.

Various simulated data samples are used in the analysis. In the simulation, $pp$
collisions are generated using 
\textsc{Pythia}~\cite{Sjostrand:2006za,*Sjostrand:2007gs} with a specific LHCb
configuration~\cite{LHCb-PROC-2010-056}. Decays of hadronic particles are
described by \textsc{EvtGen}~\cite{Lange:2001uf}. The interaction of the
generated particles with the detector, and its response, are implemented using
the \textsc{Geant4} toolkit~\cite{Allison:2006ve, *Agostinelli:2002hh} as
described in Ref.~\cite {LHCb-PROC-2011-006}.

Tagging algorithms are used to infer the initial flavor of the $B$ meson
candidate, \textit{i.e.} whether it contained a $b$ or a $\overline{b}$ quark at
production. Each algorithm provides a decision $d$ on the flavor of the $B$
meson candidate (tag), and an estimate $\eta$ of the probability for that
decision to be incorrect (mistag probability). The knowledge of the $B$ meson
production flavor is essential for this analysis, and so only candidates for
which the tagging algorithms yield a decision are considered.

One class of flavor tagging algorithms, the opposite-side tagger (OS), exploits
the dominant production mechanism of $b$ hadrons, \textit{i.e.} the production of
$b\overline{b}$ quark pairs, by reconstructing the $b$ hadron produced in
association with the signal $B$ meson. The OS tagger uses the charge of the
electron or muon from semileptonic $b$ decays, the charge of the kaon from the
$b \!\rightarrow c \!\rightarrow s$ decay chain, and the inclusive charge of
particles associated with the secondary vertex reconstructed from the $b$ hadron
decay products; further details are described in Ref.~\cite{LHCb-PAPER-2011-027}.

A major improvement in this analysis over Ref.~\cite{sin2beta:lhcb:1fb} is the
inclusion of the same-side pion tagger (SS$\pi$), which deduces the production
flavor by exploiting pions produced in the fragmentation of the $b$ quark that
produced the signal $B$ meson or in the decay of excited $B$ mesons into the
signal $B$ meson~\cite{Gronau:1992ke,LHCb-PAPER-2014-067}. Tagging pion
candidates are selected requiring charged, high momentum and high 
transverse-momentum particles that are consistent with originating from the
associated PV. Pions are identified using information from the particle
identification detectors, and the difference between the invariant mass of the
$B$ and the $B\pi^{\pm}$ pair is required to be less than
$\SI[per-mode=symbol]{1.2}{\giga\eV\per\square\clight}$. Additionally, the flight
directions of the pion and the $B$ candidate must be compatible. If multiple
pion candidates pass the selection, the one with the highest transverse momentum
is used. The mistag probability is obtained using a neural network which is
trained on simulated events and whose inputs are global event properties and
kinematic and geometric information on the pion and $B$ signal candidates.

The tagging calibration is performed in control samples of $B$ mesons whose
final state determines the $B$ flavor at decay time, by determining a linear
correction $\omega(\eta)$ that relates the estimated mistag probability $\eta$
with the mistag probability $\omega$ observed in the control sample. To account
for asymmetries in the detection efficiency of charged particles, which can lead
to different mistag probabilities for $B^0$ and ${\kern 0.18em\overline{\kern
-0.18em B}}{}^0$ mesons, an additional linear correction function
$\Delta\omega(\eta)$ is determined. Asymmetries in the efficiency of the
algorithms in determining a decision are found to be negligible.

The $B^{+} \!\rightarrow {J\mskip -3mu/\mskip -2mu\psi\mskip 2mu} K^{+}$ decay is
used to determine the flavor tagging calibration for the OS tagger. A
consistency check of the calibration is performed in a control sample of $B^0
\!\rightarrow {J\mskip -3mu/\mskip -2mu\psi\mskip 2mu} K^{*0}$ decays, showing a
good correspondence of the calibration between $B{}^{+}$ and $B{}^{0}$ decays.
As the quarks that accompany the $b$ quark in $B^{+}$ and $B^{0}$ mesons differ,
the SS$\pi$ tagger calibration is performed with $B^0 \!\rightarrow {J\mskip
-3mu/\mskip -2mu\psi\mskip 2mu} K^{*0}$ decays~\cite{LHCb-PAPER-2012-032}.
Systematic uncertainties are assigned for the uncertainties associated with the
calibration method and for the validity of the calibration in the signal decay
mode. A summary of the calibration results is given in
%Ref.~\cite{supplemental}.
\hyperref[sec:Supplementary:tagging]{the Appendix}.

The effective tagging efficiency is the product of the probability for reaching
a tagging decision, $\varepsilon_{\text{tag}} =\SI[separate-uncertainty]{36.54 \pm
0.14}{\percent}$, and the square of the effective dilution,
\mbox{$D\equiv1-2\omega = \SI[separate-uncertainty]{28.75 \pm 0.24}{\percent}$}, 
which corresponds to an effective mistag probability of $\omega = \SI
[separate-uncertainty]{35.62 \pm 0.12}{\percent}$. Compared to the previous LHCb
analysis~\cite{sin2beta:lhcb:1fb} the effective tagging efficiency
$\varepsilon_\text{eff} = \varepsilon_\text{tag} D^2$ increases from
$\SI{2.38}{\percent}$ to $\SI{3.02}{\percent}$, mainly due to the inclusion of
the SS$\pi$ tagger.

The values of the $C\!P$ violation observables $S$ and $C$ are estimated by
maximizing the likelihood of a probability density function (PDF) describing the
unbinned distributions of the following observables: the reconstructed mass $m$,
the decay time $t$ and its uncertainty estimate $\sigma_{t}$, the OS and
SS$\pi$ flavor tag decisions $d_{\text{OS}}$ and $d_{\text{SS}\pi}$, and the
corresponding per-candidate mistag probability estimates $\eta_{\text{OS}}$ and
$\eta_{\text{SS}\pi}$. The fit is performed simultaneously in $\num{24}$
independent subsamples, chosen according to data-taking conditions
($\SI{7}{\tera\eV}$, $\SI{8}{\tera\eV}$), ${K^0_{\rm\scriptscriptstyle S}}$ type
(downstream, long), flavor tagging algorithm (OS only, SS$\pi$ only, OS and
SS$\pi$), and two trigger requirements. In each category the data distribution
is modeled using a sum of two individual PDFs, one for the $B^{0}$ signal and
one for the combinatorial background.

\begin{figure}[t]
\centering
\includegraphics[width=0.48\textwidth]{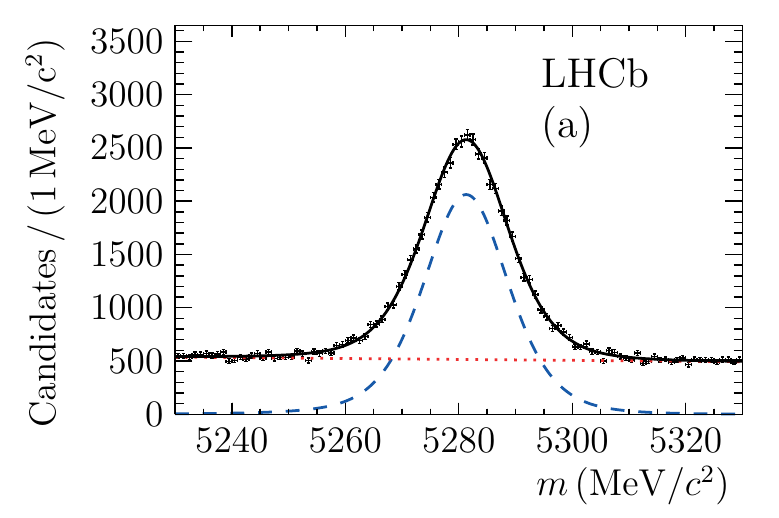}
\includegraphics[width=0.48\textwidth]{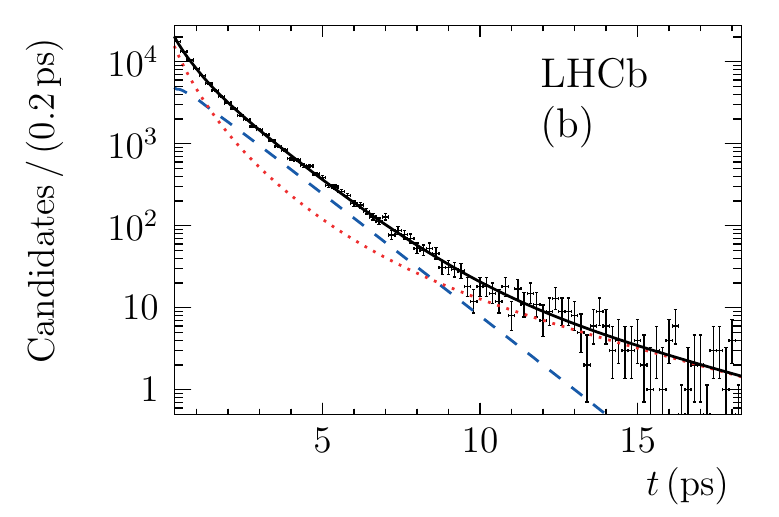}
\caption{
Distribution of (a) the reconstructed mass and (b) logarithmic distribution of
the decay time of tagged $B^0 \!\rightarrow {J\mskip -3mu/\mskip -2mu\psi\mskip
2mu} {K^0_{\rm\scriptscriptstyle S}}$ candidates. The solid black lines show the
fit projections, while the dashed (dotted) lines show the projections for the
signal (background) components only. }
\label{fig:mass_and_time}
\end{figure}

The reconstructed mass of the signal component is parametrized with a 
double-sided Hypatia PDF~\cite{Santos:2013ky} with tail parameters determined
from simulation. An exponential function is used to model the background
component, with independent parameters for the downstream and long
${K^0_{\rm\scriptscriptstyle S}}$ subsamples. The fit to the mass distributions
yields $\num[separate-uncertainty=true]{41560\pm 270}$ tagged $B^0 \!\rightarrow
{J\mskip -3mu/\mskip -2mu\psi\mskip 2mu} {K^0_{\rm\scriptscriptstyle S}}$ signal
decays. The mass distribution and projections of the PDFs are shown
in~\cref{fig:mass_and_time}\,(a).

The decay-time resolution is modeled by a sum of three Gaussian functions with
common mean, but different widths, which are convolved with the PDFs describing
the decay-time distributions. Two of the widths are given by the per-candidate
resolution estimate $\sigma_{t}$, each calibrated with independent linear
calibration functions. The third Gaussian describes the resolution for
candidates associated to a wrong PV. The scale and width parameters are obtained
in a fit to the decay-time distribution of a control sample of $B^0$ candidates
formed from prompt ${J\mskip -3mu/\mskip -2mu\psi\mskip 2mu}$ and
${K^0_{\rm\scriptscriptstyle S}}$ mesons. The parameters are determined
separately for candidates formed from downstream and long
${K^0_{\rm\scriptscriptstyle S}}$ candidates.

Trigger, reconstruction, and selection criteria distort the measured $B^{0}$
decay-time distribution, leading to a decay-time dependent efficiency. Effects
of the trigger requirements, which distort the decay-time distribution at low
decay times, are determined using data. The misreconstruction of tracks leads to
inefficiencies at large decay times. To account for this effect, an additional
decay-time dependent efficiency of the form $e^{-\beta_{t}t}$ is used, where
$\beta_{t}$ is obtained from simulation.

The PDF of true decay times $t'$ is given by
\begin{equation}
\begin{aligned}
  &\mathcal{P}\left(t',d_{\text{OS}},d_{\text{SS}\pi}\vert\eta_{\text{OS}},\eta_{\text{SS}\pi}\right) \\
  &= \sum_{d'} \left[\prod_{j} \zeta(d_j,\eta_j,d')\right]
      (1 - d' A_\text{P})\,
      e^{-t'/\tau}\left\{1 - d' S \sin(\mathrm{\Delta} m t') + d' C \cos(\mathrm{\Delta} m t')\right\},
\end{aligned}
\end{equation}
where the tag decision $d$ takes the value $+1$ ($-1$) for a tagged $B^0$
(${\kern 0.18em\overline{\kern -0.18em B}}{}^0$) candidate and $d'$ takes the
value $+1$ ($-1$) for the $B^0$ (${\kern 0.18em\overline{\kern -0.18em B}}{}^0$)
component of the signal distribution, $\tau$ is the $B^0$ meson lifetime, and
\begin{equation}
  \zeta(d_j,\eta_j,d') = 1 + d_{j}\left(1 - 2\left[\omega(\eta_{j}) 
                           + d' \frac{\Delta\omega(\eta_{j})}{2}\right]\right)
\end{equation}
represents the calibration of the tagging response from the tagging algorithm
$j=\{\text{OS},\text{SS}\pi\}$. The production asymmetry $A_\text{P}\equiv
{[\sigma({{\kern 0.18em\overline{\kern -0.18em
B}}{}^0})-\sigma({B^{0}})]}/{[\sigma({{\kern 0.18em\overline{\kern -0.18em
B}}{}^0})+\sigma({B^{0}})]}$, where $\sigma$ denotes the production 
cross-section inside the LHCb acceptance, is obtained using a measurement in
$\SI{7}{\tera\eV}$ $pp$ collisions~\cite{LHCb-PAPER-2014-042}. Considering
differences between the $7$ and $\SI{8}{\tera\eV}$ data-taking conditions, the
production asymmetries are determined as $A_\text{P}^{\SI{7}{\tera\eV}} = \num
[parse-numbers=false] {-0.0108 \pm 0.0052 \,\text{(stat)} \pm 0.0014
\,\text{(syst)}}$ and $A_\text{P}^{\SI{8}{\tera\eV}} = A_\text{P}^{\SI{7}{\tera\eV}} +
\Delta A_\text{P}$ with $\Delta A_\text{P} = \num [parse-numbers=false]{0.0004
\pm 0.0018 \,\text{(syst)}}$~\cite{LHCb-PAPER-2014-053}. The background 
decay-time distribution is parametrized by a sum of exponential functions,
convolved with the resolution model used for the signal. This parametrization
does not depend on the tag decision and mistag probability estimates. The number
of required exponential functions varies across subsamples. The decay-time
distribution and projections of the PDFs are shown in
\cref{fig:mass_and_time}\,(b). The distributions of the per-candidate resolution
estimate $\sigma_{t}$ and the per-candidate mistag probabilities,
$\eta_{\text{OS}}$ and $\eta_{\text{SS}\pi}$, are modeled by empirical
functions. Independent parameterizations are chosen for the signal and
background components.

\begin{figure}[bt]
\centering
\includegraphics[width=0.48\textwidth]{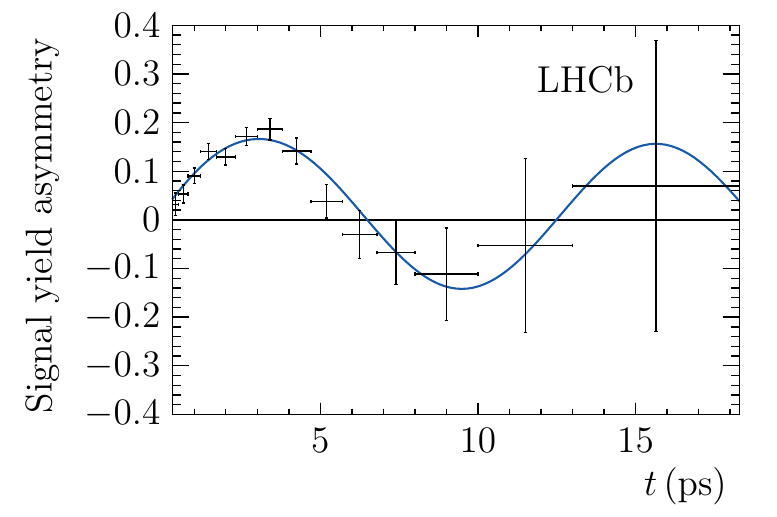}
\caption{
Time-dependent signal-yield asymmetry $(N_{{\kern 0.18em\overline{\kern -0.18em
B}}{}^0} - N_{B^0})/(N_{{\kern 0.18em\overline{\kern -0.18em B}}{}^0} +
N_{B^0})$. Here, $N_{B^0}$ ($N_{{\kern 0.18em\overline{\kern -0.18em B}}{}^0}$)
is the number of $B^0 \!\rightarrow {J\mskip -3mu/\mskip -2mu\psi\mskip 2mu}
{K^0_{\rm\scriptscriptstyle S}}$ decays with a $B^0$ (${\kern
0.18em\overline{\kern -0.18em B}}{}^0$) flavor tag. The data points are obtained
with the \textit{sPlot} technique~\cite{Pivk:2004ty}, assigning signal weights
to the events based on a fit to the reconstructed mass distribution. The solid
curve is the projection of the signal PDF.}
\label{fig:asymmetry}
\end{figure}

The likelihood is a function of $\num{83}$ free parameters, including $S$ and
$C$, and $\num{48}$ yield parameters for the signal and the background
components in $\num{24}$ individual subsamples. Eleven parameters are external
inputs, including the production asymmetry, the flavor tagging calibration
parameters, and the mass difference $\mathrm{\Delta} m$~\cite{PDG2014}. These
are constrained in the fit within their statistical uncertainties and taking
their correlations into account. The likelihood fit yields $S = \num
[parse-numbers=false]{0.729 \pm 0.035}$ and $C = \num[parse-numbers=false]
{-0.033 \pm 0.032}$ with a correlation coefficient of $\rho(S,C)=\num{0.483}$.
\Cref{fig:asymmetry} shows the decay-time dependent signal-yield asymmetry. 
An additional fit with fixed $C=\num{0}$ yields $S=\num[separate-uncertainty=true]
{0.746 \pm 0.030}$. Corrections of $\num[retain-explicit-plus]{+0.002}$ for $S$
and $\num [retain-explicit-plus]{-0.005}$ for $C$ are applied to account for
$C\!P$ violation in $K^0$--${\kern 0.2em\overline{\kern -0.2em K}}{}^0$ mixing
and for the difference in the nuclear cross-sections in material between $K^0$
and ${\kern 0.2em\overline{\kern -0.2em K}}{}^0$
states~\cite{Fetscher:1996fa,*Ko:2010mk}. The correction is negligible for the
result for $S$ with $C=0$.

Various sources of systematic uncertainties on the $C\!P$ observables are
examined, in particular from mismodeling PDFs and from systematic uncertainties
on the input parameters. In each study, a large set of pseudoexperiments is
simulated using a PDF modified such as to include the systematic effect of
interest; the relevant distributions from these pseudoexperiments are then
fitted with the nominal PDF. Significant average deviations of the fit results
from the input values are used as estimates of systematic uncertainties. The
largest systematic uncertainty on $S$, $\num{\pm 0.018}$, accounts for possible
tag asymmetries in the background; for $C$ the largest uncertainty, $\num{\pm
0.0034}$, results from the systematic uncertainty on $\mathrm{\Delta} m$.
Systematic uncertainties on the flavor tagging calibration account for the
second largest systematic uncertainty on $S$, $\num{\pm 0.006}$, and on $C$,
$\num{\pm 0.0024}$. The third largest uncertainty on $S$, $\num{\pm 0.005}$,
arises from assuming $\mathrm{\Delta} \Gamma=0$ and is evaluated by generating
pseudoexperiments with $\mathrm{\Delta} \Gamma$ set to the value of its current
uncertainty, $\SI{0.007}{\per\pico\second}$~\cite{HFAG}, and then neglecting it
in the fit. Remaining uncertainties due to neglecting correlations between the
reconstructed mass and decay time of the candidates, mismodeling of the 
decay-time resolution and efficiency, the systematic uncertainty of the
production asymmetry, and the uncertainty on the length scale of the vertex
detector are small and are given in
%Ref.~\cite{supplemental}. 
\hyperref[sec:Supplementary:systunc]{the Appendix}.
Adding all contributions in quadrature results in total systematic uncertainties
of $\num{\pm0.020}$ on $S$ and $\num{\pm0.005}$ on $C$.
 
Several consistency checks are performed by splitting the data set according to
different data-taking conditions, tagging algorithms, and different
reconstruction and trigger requirements. All results show good agreement with
the nominal results.

In conclusion, a measurement of $C\!P$ violation in the interference between the
direct decay and the decay after $B^0$--${\kern 0.18em\overline{\kern -0.18em
B}}{}^0$ oscillation to a ${J\mskip -3mu/\mskip -2mu\psi\mskip 2mu}
{K^0_{\rm\scriptscriptstyle S}}$ final state is performed using $\num{41500}$
flavor-tagged $B^0 \!\rightarrow {J\mskip -3mu/\mskip -2mu\psi\mskip 2mu}
{K^0_{\rm\scriptscriptstyle S}}$ decays reconstructed with the LHCb detector in
a sample of proton-proton collisions at center-of-mass energies of $\num{7}$ and
$\SI{8}{\tera\eV}$, corresponding to an integrated luminosity of
$\SI{3.0}{\per\femto\barn}$. The $C\!P$ observables $S$ and $C$, which allow the
determination of the CKM angle $\beta$, are measured to be
\begin{align*}
  S &= \phantom{+}0.731 \pm 0.035\,\text{(stat)} \pm 0.020\,\text{(syst)}, \\
  C &= - 0.038 \pm 0.032\,\text{(stat)} \pm 0.005\,\text{(syst)},
\end{align*} 
with a statistical correlation coefficient $\rho(S,C)=0.483$. When $C$ is fixed
to zero the measurement yields $S=\sin(2\beta)=\num [separate-uncertainty=true]
{0.746 \pm 0.030}\,\text{(stat)}$. This measurement supersedes the previous LHCb
result obtained with $\SI{1.0}{\per\femto\barn}$~\cite{sin2beta:lhcb:1fb}, and
represents the most precise time-dependent $C\!P$ violation measurement at a
hadron collider to date. Furthermore, the result has a similar precision to, and
is in good agreement with, previous measurements performed at the Belle and
BaBar experiments at the KEKB and PEP-II
colliders~\cite{sin2beta:babar,sin2beta:belle}. This result is in excellent
agreement with the expectations from other measurements and improves the
consistency of the CKM sector of the Standard Model.

%!TEX root = ../main.tex

\section*{Acknowledgements}

We express our gratitude to our colleagues in the CERN
accelerator departments for the excellent performance of the LHC. We
thank the technical and administrative staff at the LHCb
institutes. We acknowledge support from CERN and from the national
agencies: CAPES, CNPq, FAPERJ and FINEP (Brazil); NSFC (China);
CNRS/IN2P3 (France); BMBF, DFG, HGF and MPG (Germany); INFN (Italy); 
FOM and NWO (The Netherlands); MNiSW and NCN (Poland); MEN/IFA (Romania); 
MinES and FANO (Russia); MinECo (Spain); SNSF and SER (Switzerland); 
NASU (Ukraine); STFC (United Kingdom); NSF (USA).
The Tier1 computing centres are supported by IN2P3 (France), KIT and BMBF 
(Germany), INFN (Italy), NWO and SURF (The Netherlands), PIC (Spain), GridPP 
(United Kingdom).
We are indebted to the communities behind the multiple open 
source software packages on which we depend. We are also thankful for the 
computing resources and the access to software R\&D tools provided by Yandex LLC (Russia).
Individual groups or members have received support from 
EPLANET, Marie Sk\l{}odowska-Curie Actions and ERC (European Union), 
Conseil g\'{e}n\'{e}ral de Haute-Savoie, Labex ENIGMASS and OCEVU, 
R\'{e}gion Auvergne (France), RFBR (Russia), XuntaGal and GENCAT (Spain), Royal Society and Royal
Commission for the Exhibition of 1851 (United Kingdom).

\addcontentsline{toc}{section}{References}
\setboolean{inbibliography}{true}
\bibliographystyle{LHCb}
\bibliography{bibliography}

\clearpage
%!TEX root = ../main.tex

\clearpage

{\noindent\bf\Large Appendix}

\appendix

\setcounter{equation}{0}
\setcounter{table}{0}
\setcounter{figure}{0}
\renewcommand{\theequation}{A\arabic{equation}}
\renewcommand{\thetable}{A\arabic{table}}
\renewcommand{\thefigure}{A\arabic{figure}}

\subsection*{Overview of tagging calibration parameters}
\label{sec:Supplementary:tagging}

\begin{figure}[!htb]
\includegraphics[width=0.48\textwidth]{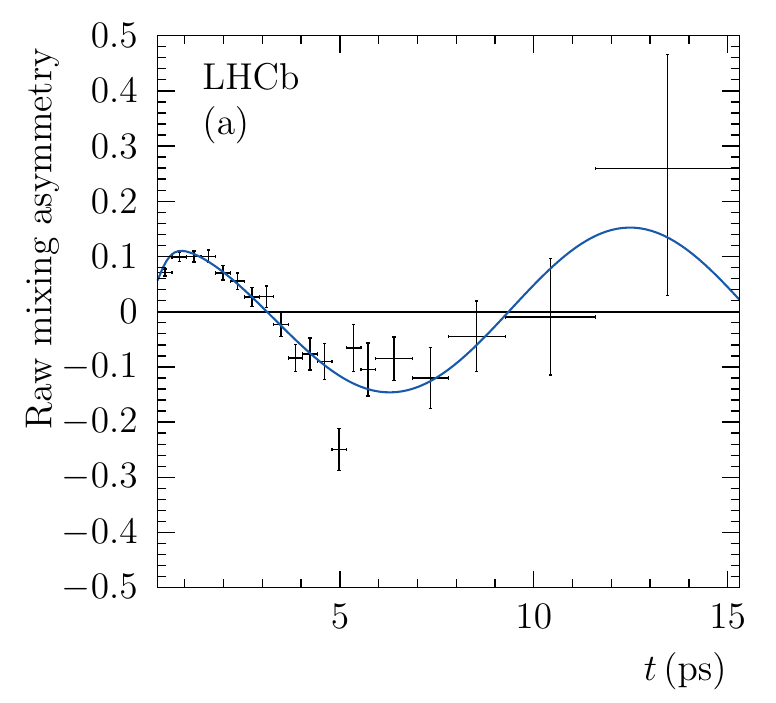}
\includegraphics[width=0.48\textwidth]{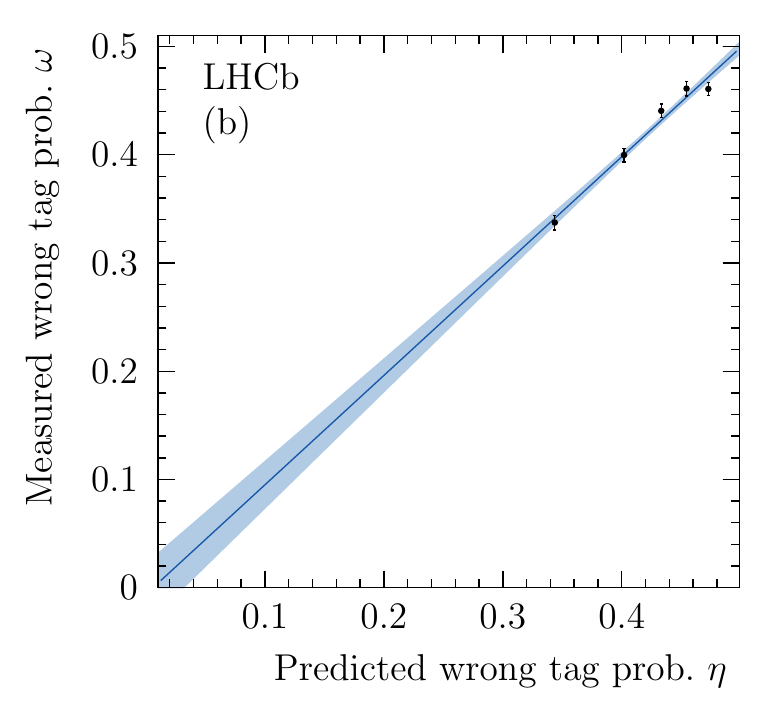}
\caption{(a) Raw mixing asymmetry $(N_{\text{unmixed}} -
N_{\text{mixed}})/(N_{\text{unmixed}} + N_{\text{mixed}})$ for all SS$\pi$
tagged $B^0\!\rightarrow {J\mskip -3mu/\mskip -2mu\psi\mskip 2mu} K^{*0}$ decay
candidates. Here, $N_{\text{unmixed}}$ ($N_{\text{mixed}}$) is the number of
$B^0\!\rightarrow {J\mskip -3mu/\mskip -2mu\psi\mskip 2mu} K^{*0}$ decays with a
final state that does (not) correspond to the flavor tag. The black line shows
the fit projection. (b) The linear calibration of the SS$\pi$ mistag probability
with $B^0\!\rightarrow {J\mskip -3mu/\mskip -2mu\psi\mskip 2mu} K^{*0}$ decays.}
\label{fig:tagging:sspion}
\end{figure}

The calibration functions of the mistag probability $\omega(\eta)$ and the
mistag probability difference $\Delta\omega(\eta)=\omega^{B^0} - \omega^{{\kern 0.18em\overline{\kern -0.18em B}}{}^0}$
are chosen as
\begin{equation}
   \omega(\eta)       = p_1 (\eta - \langle\eta\rangle) + p_0 , \qquad
   \Delta\omega(\eta) = \Delta p_1 (\eta - \langle\eta\rangle) + \Delta p_0\ .
 \end{equation} 
The OS calibration parameters are determined to be
\begin{equation}
  \begin{split}
    p_{0}^{\text{OS}}               &= 0.3815 \phantom{}   \pm 0.0011 \phantom{}  \text{\,(stat)} \pm 0.0016 \phantom{}  \text{\,(syst)} \ , \\
    p_{1}^{\text{OS}}               &= 0.978  \phantom{0}  \pm 0.012  \phantom{0} \text{\,(stat)} \pm 0.009  \phantom{0} \text{\,(syst)} \ , \\
    \Delta p_{0}^{\text{OS}}        &= 0.0148 \phantom{}   \pm 0.0016 \phantom{}  \text{\,(stat)} \pm 0.0008 \phantom{}  \text{\,(syst)} \ , \\
    \Delta p_{1}^{\text{OS}}        &= 0.070  \phantom{0}  \pm 0.018  \phantom{0} \text{\,(stat)} \pm 0.004  \phantom{0} \text{\,(syst)} \ , \\
    \langle\eta^{\text{OS}}\rangle  &= 0.3786 \ . \\
  \end{split}
\end{equation}
\Cref{fig:tagging:sspion} shows
the raw mixing asymmetry and the calibration of the mistag estimates. The
measured SS$\pi$ calibration parameters are
\begin{equation}
  \begin{split}
    p_{0}^{\text{SS}\pi}               &= \phantom{-}0.4232 \phantom{}  \pm 0.0029 \phantom{}  \text{\,(stat)} \pm 0.0028 \phantom{}   \text{\,(syst)} \ , \\
    p_{1}^{\text{SS}\pi}               &= \phantom{-}1.011  \phantom{0} \pm 0.064  \phantom{0} \text{\,(stat)} \pm 0.031  \phantom{0}  \text{\,(syst)} \ , \\
    \Delta p_{0}^{\text{SS}\pi}        &= -0.0026           \phantom{}  \pm 0.0043 \phantom{}  \text{\,(stat)} \pm 0.0027 \phantom{}   \text{\,(syst)} \ , \\
    \Delta p_{1}^{\text{SS}\pi}        &= -0.171            \phantom{0} \pm 0.096  \phantom{0} \text{\,(stat)} \pm 0.04   \phantom{00} \text{\,(syst)} \ , \\
    \langle\eta^{\text{SS}\pi}\rangle  &= \phantom{-}0.425  \ . \\
  \end{split}
\end{equation}

In the dataset of 114000 $B^0 \!\rightarrow {J\mskip -3mu/\mskip -2mu\psi\mskip
2mu} {K^0_{\rm\scriptscriptstyle S}}$ decays, the OS tagger algorithm yields a
tagging power of $(2.63\pm0.04)\%$ and the SS$\pi$ tagger algorithm yields
$(0.376\pm0.024)\%$. For events that have both OS and SS$\pi$ tagging decisions,
the effective tagging power is $(0.503 \pm 0.010)\%$. The combined effective
tagging power of these three overlapping tagging categories is $(3.02 \pm
0.05)\%$.

\subsection*{Summary of systematic uncertainties}
\label{sec:Supplementary:systunc}

The systematic uncertainties are summarised in \cref{tab:syst_total}. The
overall systematic uncertainty is calculated by summing the single uncertainties
in quadrature. The relative systematic uncertainties compared to the central
values of $S$ and $C$ are given in brackets. Here, we set $S = \num{0.729}$ and
$C = \num{-0.033}$ as reference.
\begin{table}[!htb]
  \caption{Systematic uncertainties $\sigma_{S}$ and $\sigma_{C}$ on $S$ and
  $C$. Entries marked with a dash represent studies where no significant effect
  is observed.}
  \label{tab:syst_total}
  \centering
  \resizebox{0.95\textwidth}{!}{
    \begin{tabular}{lrlrl}
      \toprule
      Origin & \multicolumn{2}{c}{$\sigma_{S}$} & \multicolumn{2}{c}{$\sigma_{C}$}    \\
      \midrule
      \small{Background tagging asymmetry}           &  $\tablenum[table-format = 1.4]{0.0179}$ & ($\tablenum[table-format = 1.1]{2.5}\si{\percent}$) &  $\tablenum[table-format = 1.4]{0.0015}$ & ($\tablenum[table-format = 2.1]{4.5}\si{\percent}$) \\
      \small{Tagging calibration}                    &  $\tablenum[table-format = 1.4]{0.0062}$ & ($\tablenum[table-format = 1.1]{0.9}\si{\percent}$) &  $\tablenum[table-format = 1.4]{0.0024}$ & ($\tablenum[table-format = 2.1]{7.2}\si{\percent}$) \\
      \small{$\Delta \Gamma$}                                  &  $\tablenum[table-format = 1.4]{0.0047}$ & ($\tablenum[table-format = 1.1]{0.6}\si{\percent}$) &  \multicolumn{2}{c}{---}               \\
      \small{Fraction of wrong PV component}         &  $\tablenum[table-format = 1.4]{0.0021}$ & ($\tablenum[table-format = 1.1]{0.3}\si{\percent}$) &  $\tablenum[table-format = 1.4]{0.0011}$ & ($\tablenum[table-format = 2.1]{3.3}\si{\percent}$) \\
      \small{$z$-scale}                              &  $\tablenum[table-format = 1.4]{0.0012}$ & ($\tablenum[table-format = 1.1]{0.2}\si{\percent}$) &  $\tablenum[table-format = 1.4]{0.0023}$ & ($\tablenum[table-format = 2.1]{7.0}\si{\percent}$) \\
      \small{$\mathrm{\Delta} m$}                    &  \multicolumn{2}{c}{---}               &  $\tablenum[table-format = 1.4]{0.0034}$ & ($\tablenum[table-format = 2.1]{10.3}\si{\percent}$)\\
      \small{Upper decay time acceptance}            &  \multicolumn{2}{c}{---}               &  $\tablenum[table-format = 1.4]{0.0012}$ & ($\tablenum[table-format = 2.1]{3.6}\si{\percent}$) \\
      \small{Correlation between mass and decay time}&  \multicolumn{2}{c}{---}               &  \multicolumn{2}{c}{---}               \\
      \small{Decay time resolution calibration}      &  \multicolumn{2}{c}{---}               &  \multicolumn{2}{c}{---}               \\
      \small{Decay time resolution offset}           &  \multicolumn{2}{c}{---}               &  \multicolumn{2}{c}{---}               \\
      \small{Low decay time acceptance}              &  \multicolumn{2}{c}{---}               &  \multicolumn{2}{c}{---}               \\
      \small{Production asymmetry}                   &  \multicolumn{2}{c}{---}               &  \multicolumn{2}{c}{---}               \\
      \midrule
      Sum                                            &  $0.020$        & ($\tablenum[table-format = 1.1]{2.7}\si{\percent}$) & $0.005$          & ($\tablenum[table-format = 2.1]{15.2}\si{\percent}$)\\
      \bottomrule
    \end{tabular}
  }
\end{table}

\clearpage
%%%%%%%%%%%%%%%%%%%%%%%%%%%%%%%%%%%%%%%%%%
\centerline{\large\bf LHCb collaboration}
\begin{flushleft}
\small
R.~Aaij$^{41}$, 
B.~Adeva$^{37}$, 
M.~Adinolfi$^{46}$, 
A.~Affolder$^{52}$, 
Z.~Ajaltouni$^{5}$, 
S.~Akar$^{6}$, 
J.~Albrecht$^{9}$, 
F.~Alessio$^{38}$, 
M.~Alexander$^{51}$, 
S.~Ali$^{41}$
% ...

{\footnotesize \it
% ...
$ ^{5}$Clermont Universit\'{e}, Universit\'{e} Blaise Pascal, CNRS/IN2P3, LPC, Clermont-Ferrand, France\\
$ ^{6}$CPPM, Aix-Marseille Universit\'{e}, CNRS/IN2P3, Marseille, France\\
% ...
$ ^{9}$Fakult\"{a}t Physik, Technische Universit\"{a}t Dortmund, Dortmund, Germany\\
% ...
$ ^{37}$Universidad de Santiago de Compostela, Santiago de Compostela, Spain\\
$ ^{38}$European Organization for Nuclear Research (CERN), Geneva, Switzerland\\
% ...
$ ^{41}$Nikhef National Institute for Subatomic Physics, Amsterdam, The Netherlands\\
% ...
$ ^{46}$H.H. Wills Physics Laboratory, University of Bristol, Bristol, United Kingdom\\
% ...
$ ^{51}$School of Physics and Astronomy, University of Glasgow, Glasgow, United Kingdom\\
$ ^{52}$Oliver Lodge Laboratory, University of Liverpool, Liverpool, United Kingdom\\
% ...
}
\end{flushleft}
%%%%%%%%%%%%%%%%%%%%%%%%%%%%%%%%%%%%%%%%%%

\end{document}